\documentclass[preprint,showpacs,preprintnumbers,amsmath]{revtex4}
\usepackage{txfonts}
\usepackage{graphicx}
\usepackage{dcolumn}
\usepackage{bm}
\usepackage{subfigure}

\begin{document}

\title{Sign change of surface energy and stability of composite vortex in
two-component superconductivity}
\author{Jun-Ping Wang}
\affiliation{Department of Physics, Yantai University, Yantai
264005, P. R. China}

\begin{abstract}
Existence of thermodynamically stable composite vortices in a
two-component superconductor may form distinctive vortex patterns,
and may lead to type-1.5 superconductivity. Here we study the
surface energy of the two-component superconductor and show that the
sign of surface energy is determined not only by the Ginzburg-Landau
parameters $\kappa_i\;(i=1,2)$ of two superconducting components,
but also by a temperature independent parameter $\kappa _\xi $,
which is defined as the ratio of the coherence lengths of two
components. Since the negative surface energy conduces to the
invasion of thermodynamically stable composite vortices into a
superconductor, the criterions for stability of composite vortex are
these three independent dimensionless parameters. We find that there
can exist thermodynamically stable composite vortex in a
type-1+type-2 or type-2+type-2 material. We also predict that
unusual vortex patterns like those observed in $MgB_2$ (V.
Moshchalkov et al, Phys. Rev. Lett., 102, 117001 (2009)) can occur
in some type-2+type-2 superconductors.
\end{abstract}


\maketitle


It is well known that a conventional superconductor can be categorized as
type-1 or type-2, depending on its behavior under a magnetic field. In a
type-1 superconductor, such as lead or aluminum, superconductivity loss
happens suddenly as the field surpasses a critical level. A type-2
superconductor, such as niobium, has two critical fields, and if the applied
field is stronger than the lower but weaker than the higher, then the field
can penetrate the material by vortices \cite{www}. The criterion that
determines whether a superconductor is of type-1 or type-2 is the
Ginzburg-Landau (GL) parameter $\kappa $ \cite{ginzburg}. It is defined as
the ratio of the penetration depth $\lambda $ over the coherence length $\xi
$, $\kappa =\lambda /\xi $, and the critical value $\kappa _c=1/\sqrt{2}$
represents the demarcation line between type-1 $\left( \kappa <\kappa
_c\right) $ and type-2 $\left( \kappa >\kappa _c\right) $ superconductors.
Interaction between vortices is attractive in a type-1 superconductor \cite
{a1,a2} and repulsive in a type-2 superconductor \cite{abrikosov}.

Novel phenomena may be found in the newly discovered two-component
superconductors, such as $MgB_2$ \cite{MgB2}. In the experiment by
Moshchalkov et al, unconventional stripe- and gossamerlike vortex
patterns have been directly visualized by Bitter decorations on high
quality $MgB_2$ single crystals \cite{1.5}. These observations are
attributed to the coexistence of $\pi $ and $\sigma $
superconducting components in $MgB_2$. These two components are in
different regimes: $\kappa _\pi =\lambda _\pi /\xi _\pi
=0.66<1/\sqrt{2}$ (type-1) and $\kappa _\sigma =\lambda _\sigma /\xi
_\sigma =3.68>1/\sqrt{2}$ (type-2). There are evidences of existence
of composite vortex, which is axisymmetric, and has a normal core
with phase of each condensate changes $2\pi$ around the core.
Interaction between these composite vortices is short-range
repulsive and long-range attractive. This is different from that of
type-1 or type-2 superconductivity, and leads to type-1.5
superconductivity \cite{1.5,babaev}.

The key features of type-1.5 superconductivity are: (i) existence of
thermodynamically stable composite vortex; (ii) interaction between
composite vortices is short-range repulsive and long-range
attractive. It has been shown in numerical simulations that the
composite vortex in a type-1+type-2 superconductor is
thermodynamically stable when the disparity between coherence
lengths $\xi _1$ and $\xi _2$ is extremely large \cite {babaev}.
However we need definite criterion for stability of composite vortex
in a two-component superconductor.

Motivated by recent interest in exotic vortex state in
multi-component superconductors, here we study one-dimensional
superconducting-normal boundary in an abstract two-component system.
We show that, the sign of surface energy in such system is
determined not only by the Ginzburg-Landau parameters
$\kappa_i\;(i=1,2)$ of two superconducting components, but also by a
temperature independent parameter $\kappa _\xi$, which is defined as
the ratio of the coherence lengths of two components: $\kappa _\xi=
{\xi_1}/{\xi_2} $. Since the negative surface energy conduces to the
invasion of thermodynamically stable composite vortices into a
superconductor, the criterions for stability of composite vortex are
these three independent dimensionless parameters. The same method
has been used to speculate the stability of vortex in conventional
one-component superconductor \cite{abrikosov}. The vortex discussed
in present work is axisymmetric composite vortex with normal core,
and phase of each condensate changes $2\pi$ around the core. Of
course, in principle, there can exist other kinds of composite
vortices unlike those we discussed here. But it was shown that the
composite vortex discussed in current work is energetically
preferred in weak external field \cite{1.5,babaev,babaev2}. The
invasion of composite vortex is thermodynamically favorable for
certain values of $\kappa _1,\kappa _2,\kappa _\xi $, which ensure
the negative surface energy. We find that there can exist
thermodynamically stable composite vortex in a type-1+type-2 or
type-2+type-2 material.

We start with a general system in which two superconducting components
coexist. The GL free energy density of the system is
\begin{equation}
f_s=f_{n0}+\sum_{i=1}^2{\frac{\hbar ^2}{2m_i^{*}}|(\nabla -\frac{ie_i^{*}}{{%
\hbar }c}\mathbf{A})\Psi _i|^2}+V({|\Psi _{1,2}|}^2)+\eta (\Psi _1^{*}\Psi
_2+\Psi _1\Psi _2^{*})+\frac 1{8\pi }(\nabla \times \mathbf{A})^2,
\label{GL}
\end{equation}
where $f_{n0}$ is the free energy density of the body in the normal state in
the absence of the magnetic field, $V({|\Psi _i|}^2)=a_i{|\Psi _i|}^2+b_i{%
|\Psi _i|}^4/2\;(i=1,2)$. $\eta $ is a coefficient characterizes Josephson
coupling between two superconducting components. In the following we do not
consider coupling effect and set $\eta =0$. We also assume that the
effective mass $m_i^{*}$ and charge $e_i^{*}$ of two components are equal: $%
m_i^{*}=m^{*}$, $e_i^{*}=e^{*}$. There are four characteristic lengths: the penetration depth $%
\lambda _i$ and coherence length $\xi _i$ for each component are given by: $%
\lambda _i=(m^{*}c^2/4{\pi }e^{*2}\Psi _{i0}^2)^{1/2}$, $\xi _i=\hbar
/(2m^{*}|a_i|)^{1/2}$, where $\Psi _{i0}=(-a_i/b_i)^{1/2}$. The
thermodynamic critical magnetic field of the individual component is $%
H_{ct(i)}=\Phi _0/(2\sqrt{2}\pi \lambda _i\xi _i),$ where $\Phi _0=hc/e^{*}$
is the flux quantum. The magnetic field penetration depth and the
thermodynamic critical magnetic field of the system (\ref{GL}) are: $\lambda
=(1/\lambda _1^2+1/\lambda
_2^2)^{-1/2},\;H_{ct}=(H_{ct(1)}^2+H_{ct(2)}^2)^{1/2}$. Notice that $\lambda
<\min (\lambda _1,\lambda _2)$, $H_{ct}>\max (H_{ct(1)},H_{ct(2)})$.

The problem of the thermodynamical stability of the composite vortex with
normal core can be converted into the surface energy problem. The stability
of the composite vortex depends upon the surface energy, or, more
accurately, upon the sign of surface energy. Let us consider a plane
interface between normal ($n$) and superconducting ($s$) phases in a
two-component superconductor, taking the interface as the $yz$-plane and the
$x$-axis into the $s$ phase. Surface energy $\alpha _{ns}$ is defined as,
under the thermodynamic critical magnetic field $\mathbf{H}_{ct}$ =$H_{ct}%
\hat{z}$, the Gibbs energy difference between the $n$, $s$ transitional
state and the fully normal state (or fully superconducting state since these
must be equal) of the superconductor with unit cross-section:
\begin{equation}
\alpha _{ns}=\int_{-\infty }^\infty dx\left\{ {\frac{\hbar ^2}{2m^{*}}}%
\sum_{i=1}^2{|(\nabla -\frac{ie^{*}}{{\hbar }c}\mathbf{A})\Psi _i|^2}+V({%
|\Psi _{1,2}|}^2){+}\frac 1{8\pi }\left( \mathbf{H}_{ct}-\nabla \times {{%
\mathbf{A}}}\right) ^2\right\} .  \label{SE}
\end{equation}
The integrand vanishes, both within the $n$ phase $(x\rightarrow -\infty )$,
where ${\Psi _i=0}$ and $\nabla \times {{\mathbf{A=}}}\mathbf{H}_{ct}$, and
within the $s$ phase $(x\rightarrow \infty )$, where ${{\Psi _i=\Psi _{i0}}}$
and $\nabla \times {{\mathbf{A=}}}0$. Now the distribution of all quantities
depends only on the coordinate $x$. This fact enable us to choose gauge
potential as $\mathbf{A=(}0,A_y(x),0\mathbf{)}$. Then the order parameters ${%
\Psi _i}$ can be taken real. We shall use the dimensionless quantities as
following: $\rho \equiv x/\lambda ,\;\psi _1\equiv \Psi _1/\Psi _{10},\;\psi
_2\equiv \Psi _2/\Psi _{20},\;A\equiv \left| \mathbf{A}\right|
/H_{ct}\lambda ,\;A^{^{\prime }}=B\equiv \left| \nabla \times {{\mathbf{A}}}%
\right| /H_{ct}.$ Then the expression (\ref{SE}) becomes $\alpha
_{ns}=(H_{ct}^2\lambda /8\pi )\int_{-\infty }^\infty d\rho \left\{
\sum_{i=1}^2{(H_{ct(i)}/H_{ct})^2}\left[ {2(\xi _i/\lambda )^2\psi
_i^{^{\prime }2}+(H_{ct}/H_{ct(i)})^2(\lambda /\lambda _i)^2}A{^2\psi
_i^2-2\psi _i^2+\psi _i^4}\right] +(A^{^{\prime }}-1)^2\right\} .$ It can be
verified that all coefficients in the integrand can be represented as
functions of three dimensionless temperature independent parameters: $\kappa
_1\equiv \lambda _1/\xi _1,\;\kappa _2\equiv \lambda _2/\xi _2,\;\kappa _\xi
\equiv \xi _1/\xi _2$ \cite{3p}. And the surface energy (\ref{SE}) can be
rewritten as:
\begin{equation}
\alpha _{ns}=\frac{H_{ct}^2\lambda }{8\pi }\int_{-\infty }^\infty
d\rho\left\{ \sum_{i=1}^2\frac{{C_i}}{B_i}\left[ {2A}_i{\psi _i^{^{\prime
}2}+}\left( {B}_iA{^2-2}\right) {\psi _i^2+\psi _i^4}\right] +(A^{^{\prime
}}-1)^2\right\} ,  \label{SE3}
\end{equation}
where $A_1=1/\kappa _1^2+\kappa _\xi ^2/\kappa _2^2,\;B_1=(\kappa
_2^2+\kappa _1^2\kappa _\xi ^4)/(\kappa _2^2+\kappa _1^2\kappa _\xi
^2),\;C_1=\kappa _2^2/(\kappa _2^2+\kappa _1^2\kappa _\xi ^2),\;A_2=1/\kappa
_2^2+1/\kappa _1^2\kappa _\xi ^2,\;\;B_2=(\kappa _2^2+\kappa _1^2\kappa _\xi
^4)/[\kappa _\xi ^2(\kappa _2^2+\kappa _1^2\kappa _\xi ^2)],\;C_2=\kappa
_1^2\kappa _\xi ^2/(\kappa _2^2+\kappa _1^2\kappa _\xi ^2)$. The GL eqs of
motion following from the free energy (\ref{GL}) are:
\[
A_1\psi _1^{^{\prime \prime }}=\frac{B_1}2A^2\psi _1-\psi _1+\psi _1^3,
\]
\[
A_2\psi _2^{^{\prime \prime }}=\frac{B_2}2A^2\psi _2-\psi _2+\psi _2^3,
\]
\begin{equation}
A^{^{\prime \prime }}=(C_1\psi _1^2+C_2\psi _2^2)A,  \label{GL3}
\end{equation}
with boundary conditions: ${\psi }_1(-\infty )=0,\;{\psi }_2(-\infty
)=0,\;A^{^{\prime }}(-\infty )=1,\;{\psi }_1(\infty )=1,\;{\psi }_2(\infty
)=1,\;A^{^{\prime }}(\infty )=0$. The surface energy $\alpha _{ns}$ is
obtained from the substitution of field variables $\psi _1,\;\psi _2,\;A$
that satisfy the GL eqs (\ref{GL3}) into (\ref{SE3}).

It is clear from (\ref{SE3}) and (\ref{GL3}) that the sign of the surface
energy is determined by three independent dimensionless parameters: $\kappa
_1,\;\kappa _2,\;\kappa _\xi $. If these three parameters are known for a
material considered, we can then obtain the value of $\alpha
_{ns}/(H_{ct}^2\lambda /8\pi )$ from the substitution of $\psi _1,\;\psi
_2,\;A$ satisfy (\ref{GL3}) into (\ref{SE3}) and identify the sign of
surface energy. If the sign of surface energy is negative, i.e., $\alpha
_{ns}/(H_{ct}^2\lambda /8\pi )<0$, occurrence of composite vortex is
thermodynamically favorable. Otherwise, no stable composite vortex can
emerge.

When the coherence lengths of two components are equal: $\xi _1=\xi _2$,
i.e., $\kappa _\xi =1$, it is easily verified that equations (\ref{GL3})
have the first integral:
\begin{equation}
\frac{\kappa _2^2}{\kappa _1^2+\kappa _2^2}\left[ 2\left( \frac 1{\kappa
_1^2}+\frac 1{\kappa _2^2}\right) \psi _1^{^{\prime }2}+\left( 2-A^2\right)
\psi _1^2-\psi _1^4\right] +\frac{\kappa _1^2}{\kappa _1^2+\kappa _2^2}%
\left[ 2\left( \frac 1{\kappa _1^2}+\frac 1{\kappa _2^2}\right) \psi
_2^{^{\prime }2}+\left( 2-A^2\right) \psi _2^2-\psi _2^4\right] +A^{^{\prime
}2}-1=0.  \label{integral}
\end{equation}
With (\ref{integral}), the expression (\ref{SE3}) becomes $\alpha _{ns}=%
\frac{H_{ct}^2\lambda }{4\pi }\int_{-\infty }^{+\infty }d\rho \left\{ \frac
2{\kappa _1^2}\psi _1^{^{\prime }2}+\frac 2{\kappa _2^2}\psi _2^{^{\prime
}2}+A^{^{\prime }}\left( A^{^{\prime }}-1\right) \right\} $. The sign of
surface energy, thus the stability of composite vortex, can be determined by
the procedures described above. Let us comment on several particular cases:
In the case $\kappa _1\gg 1$, $\kappa _2\gg 1$, the first two terms in the
integrand can be neglected and the sign of surface energy is always negative
since $A^{^{\prime }}\in [0,1]$. The sign of surface energy is positive in
the opposite case $\kappa _1\ll 1$, $\kappa _2\ll 1$. If one component is of
extreme type-2, while the other component is of extreme type-1, i.e., $%
\kappa _1\gg 1$, $\kappa _2\ll 1$. The contribution of the first term in the
integrand can be neglected and the sign of surface energy is positive.

Let us now study general cases in which there is disparity in coherence
lengths between two components: $\xi _1\neq \xi _2$, i.e., $\kappa _\xi \neq
1$. We then need to solve eqs (\ref{GL3}) with given boundary conditions
numerically . Before a detailed numerical work is undertaken, we first
analyze the problem qualitatively. We note that in the integrand in (\ref{SE}%
) only the second term $V({|\Psi _{1,2}|}^2)$ contributes the negative value
to surface energy. The distance scale over which the condensates tends to
its expectation value ${\Psi _{i0}}$ is of order $\sim \;\xi _i$. $V({|\Psi
_i|}^2)$ decreases from $0$ to $a_i{|\Psi _{i0}|}^2+b_i{|\Psi _{i0}|}%
^4/2=-H_{ct(i)}^2/(8\pi )$ in the same range. The length scale over which
the magnetic field decays is $\sim \lambda $. The last term of integrand in (%
\ref{SE}) $\frac{1\;}{8\pi }\left( \mathbf{H}_{ct}-\nabla \times {{\mathbf{A}%
}}\right) ^2$ increases from $0$ to $\frac{1\;}{8\pi }H_{ct}^2$ in this
range. The similar term in the surface energy expression when only single
superconducting component exists increases from $0$ to $\frac{1\;}{8\pi }%
H_{ct(i)}^2$ in a range $\sim \lambda _i$. Since $\lambda <\min (\lambda
_1,\lambda _2)$, $H_{ct}=(H_{ct(1)}^2+H_{ct(2)}^2)^{1/2},$ the integral
value of the last term in integrand in (\ref{SE}) is positive and is much
larger than the sum of the integral values of similar terms in the surface
energy expressions when single component exists. We then conclude that there
is a trend of increase in surface energy for a two-component superconductor
comparing to the sum of surface energy of the single component cases. The
detailed numerical simulations below confirm this idea.

To explore the concrete behavior of the sign of surface energy with three
parameters $\kappa _1,\;\kappa _2,\;\kappa _\xi $, we then search for the
numerical solutions of the GL eqs. (\ref{GL3}). And we really identified the
sign change of the surface energy due to the variation of these three
parameters. There are three cases:

\emph{Case 1.} $\kappa _1<1/\sqrt{2}$, $\kappa _2<1/\sqrt{2}$, i.e., two
components are both of type-1. As we have shown, there is a trend of
increase in surface energy for a two-component superconductor comparing to
the sum of surface energy of the single component cases. Since the sign of
surface energy for a type-1 material is always positive, we then conclude
that the system has positive surface energy and there is no
thermodynamically stable composite vortex.

\begin{figure}[tbp]
\centering
\includegraphics[scale=0.8,trim=0 0 0 0]{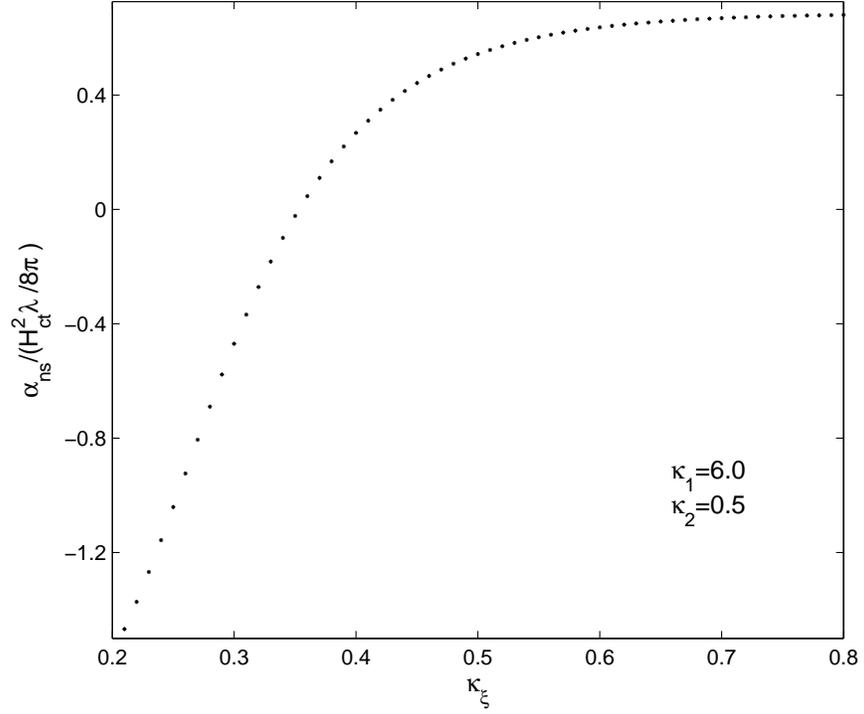}
\caption{Sign change of surface energy for a two component superconductor
with $\kappa _1=6.0,\;\kappa _2=0.5$. It is shown that the sign of surface
energy changes from positive to negative with decreasing parameter $\kappa
_\xi $. Note that the negative surface energy conduces to the invasion of
thermodynamically stable vortex. The critical value of $\kappa _\xi $ at
which surface energy vanishes is $\kappa _{\xi c}=0.35$.}
\label{fig:figu1}
\end{figure}

\emph{Case 2.} $\kappa _1>1/\sqrt{2}$, $\kappa _2<1/\sqrt{2}$, i.e.,
the first component is of type-2, while the second is of type-1.
When the third parameter $\kappa _\xi \gg 1$, i.e., $\xi _1\gg \xi
_2$, thus $\lambda _1>\xi _1/\sqrt{2}\gg \xi _2>\lambda _2$,
$\lambda =(1/\lambda _1^2+1/\lambda _2^2)^{-1/2}\approx \lambda
_2<\xi _2$. Then the integral value of the last term in integrand in
(\ref{SE}) $\frac{1\;}{8\pi }\left( \mathbf{H}_{ct}-\nabla \times
{{\mathbf{A}}}\right) ^2$ gains ascendancy over that of the second
term $V({|\Psi _{1,2}|}^2)$, and the sign of surface energy tends to
be positive. On the other hand, when $\kappa _\xi \ll 1$ , i.e.,
$\xi _1\ll \xi _2$, penetration depth $\lambda $ may fall into the
region $\xi _1<\lambda <\xi _2$, and the sign of the surface energy
can take negative. We then conclude that there is a critical value
$\kappa _{\xi c}$ at which the surface energy vanishes. As an
example, we show in Fig. \ref {fig:figu1} the sign change of surface
energy for a two-component superconductor with $\kappa _1=6.0$,
$\kappa _2=0.5$. It is clear that the sign of surface energy is
negative when $\kappa _\xi <\kappa _{\xi c}=0.35$. Generally, for
fixed $\kappa _1>1/\sqrt{2}$, $\kappa _2<1/\sqrt{2}$, the critical
value $\kappa _{\xi c}$ can be determined using numerical method as
shown above. However, a rough estimate of the upper limit of the
critical value $\kappa _{\xi c} $ can be made based on the condition
$\xi _1<\lambda <\xi _2$, under which the sign of surface energy can
be expected to be negative,
\begin{equation}
\kappa _{\xi c}<\kappa _2\sqrt{1-\frac 1{\kappa _1^2}}.
\label{kappac}
\end{equation}
Note that the negative surface energy conduces to the invasion of
thermodynamically stable composite vortices in a two-component
material, and these composite vortices assume full responsibility
for type-1.5 superconductivity in a type-1+type-2 superconductor.
Then we can regard Eq. (\ref{kappac}) as a rough criterion for
type-1.5 superconductivity in the two-component superconductor in
which type-1 and type-2 condensates coexist.

\begin{figure}[tbp]
\begin{center}
\includegraphics[scale=0.8,trim=0 0 0 0]{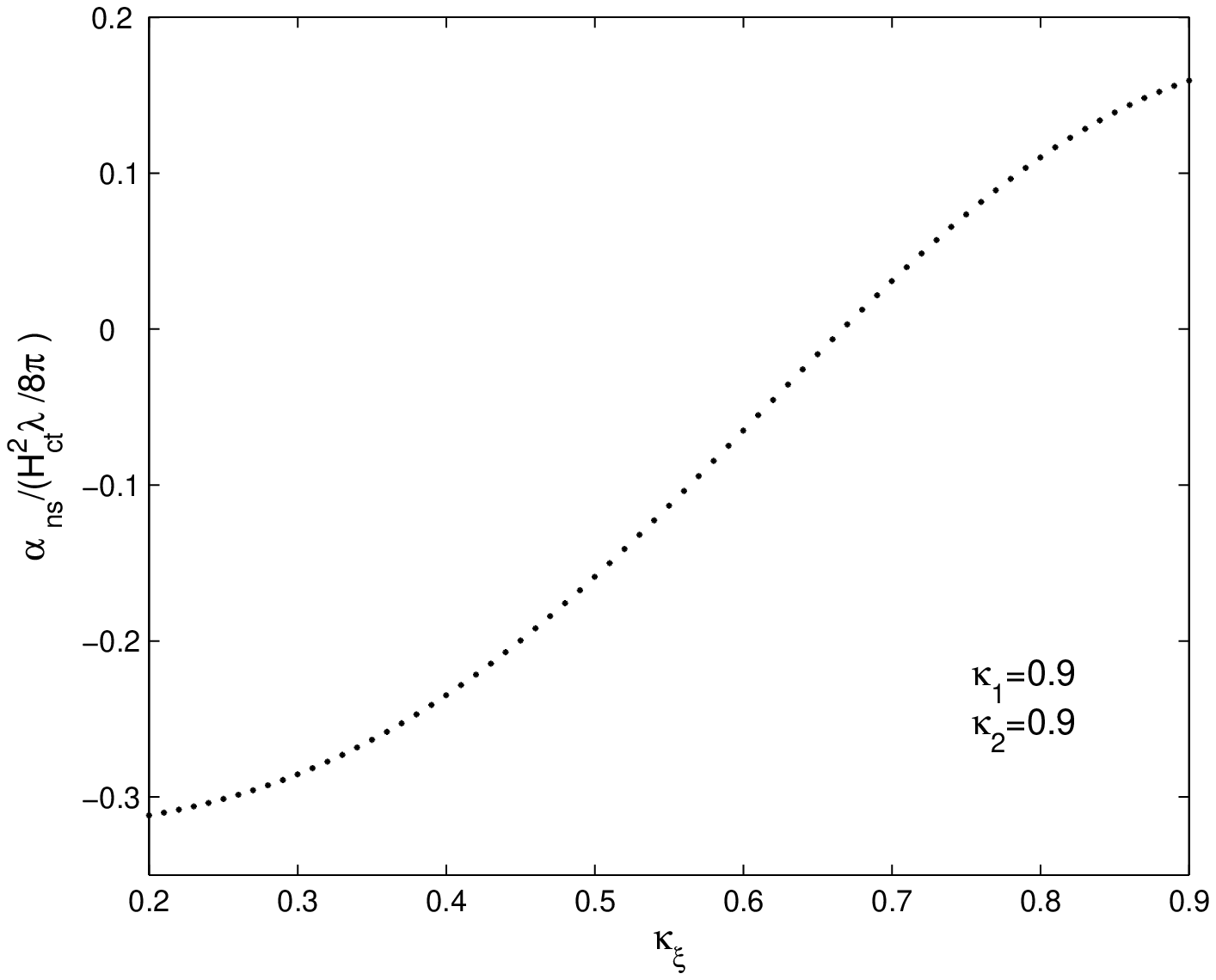}
\end{center}
\caption{Sign change of surface energy for a two component superconductor
with $\kappa _1=\kappa _2=0.9$. The critical value of $\kappa _\xi $ at
which surface energy vanishes is $\kappa _{\xi c}=0.67$.}
\label{fig:figu2}
\end{figure}

\emph{Case 3.} $\kappa _1>1/\sqrt{2}$, $\kappa _2>1/\sqrt{2}$, i.e.,
two components are both of type-2. The sign of surface energy of the
system is then expected to be negative. However, when $\kappa
_1\approx \kappa _2\approx 1$, $\kappa _\xi \approx 1$, the effect
of coexistence of the two superconducting components may result in a
positive surface energy. In Fig. \ref{fig:figu2} we show the
numerical result in the case $\kappa _1=\kappa _2=0.9$. It is clear
to see that there is a sign change of the surface energy from
negative to positive with increasing parameter $\kappa _\xi $. This
shows the the condition conversion for existence and nonexistence of
composite vortex.

At last, we discuss the possible vortex patterns which will occur in
a type-2+type-2 superconductor after the stability of composite
vortex has been confirmed. We concentrate on two cases: (i) if the
penetration depth is much larger than the coherence lengths of both
condensates, $\lambda\gg \xi_{i}\;(i=1,2)$, two vortices will have
their supercurrents overlapping first and repel each other. Then,
regular vortex patterns like those in an usual type-2 superconductor
will occur. (ii) if the disparity between the coherence lengths of
two components is large, and the penetration depth is much smaller
than the coherence length of one component, i.e., $\lambda\ll \xi_1$
(or $\lambda\ll \xi_2$), the vortex has an extended core associated
with the condensate $\Psi_1$ (or $\Psi_2$). Interaction between two
vortices is expected to be short-range repulsive and long-range
attractive, similar to that of stable composite vortices in the
type-1+type-2 case. We then expect unusual vortex patterns like
those observed in $MgB_2$ \cite{1.5} will occur in these
type-2+type-2 materials.

In conclusion, coexistence of two superconducting components in a material
may lead to the occurrence of stable composite vortex and distinctive
response of material to applied magnetic field. We have shown that the
criterions for stability of composite vortex with normal core in a two
component superconductor are three independent dimensionless parameters: $%
\kappa _1,\;\kappa _2,\;\kappa _\xi $. And there can exist stable composite
vortex in a type-1+type-2 or type-2+type-2 superconductor.

This work was supported by the National Natural Science Foundation of China
(No. 10547137).

\begin{references}
\bibitem{3p} In fact here we need at least three parameters to express ratios of four characteristic
lengths: $\lambda_{1},\;\lambda_{2},\;\xi_{1},\;\xi_{2}$. We select
$\kappa_1,\;\kappa _2,\;\kappa _\xi$ because they have definite
physical meaning and help to give an intuitive estimate of the sign
of the surface energy.
\end{references}
\end{document}